\documentclass[aps,floatfix,onecolumn,nofootinbib]{revtex4}
\usepackage{graphicx}
\usepackage{graphics,epsfig}
\usepackage{amsmath,amssymb,amsfonts}

\usepackage{color}
\definecolor{red}{rgb}{1,0,0}

\begin{document}
\title{Singularities in loop quantum cosmology}
\date{October 20, 2008}

\author{Thomas Cailleteau $^{1,2}$}\email{thomas.cailleteau@gmail.com}
\author{Antonio Cardoso $^2$}\email{antonio.cardoso@port.ac.uk}
\author{Kevin Vandersloot $^2$}\email{kevin.vandersloot@port.ac.uk}
\author{David Wands $^2$}\email{david.wands@port.ac.uk}
\affiliation{$^1$ Physique Fondamentale et appliqu\'ee,
Universite Paris-Sud 11, 91405 Orsay Cedex, France \\
$^2$ Institute of Cosmology and Gravitation, University of
Portsmouth, Portsmouth P01 2EG, United Kingdom}

\begin{abstract}

We show that simple scalar field models can give rise to curvature
singularities in the effective Friedmann dynamics of Loop Quantum
Cosmology (LQC). We find singular solutions for spatially flat
Friedmann-Robertson-Walker cosmologies with a canonical scalar field
and a negative exponential potential, or with a phantom scalar field
and a positive potential. While LQC avoids big bang or big rip type
singularities, we find sudden singularities where the Hubble rate is
bounded, but the Ricci curvature scalar diverges. We conclude that
the effective equations of LQC are not in themselves sufficient to
avoid the occurrence of
curvature singularities.

\end{abstract}
\maketitle

The absence of singularities may be taken to be a pre-requisite for
any fundamental theory of nature, and in particular we expect a
quantum theory of gravity to resolve the curvature singularities
found in Einstein's theory of general relativity.
Loop quantum cosmology (LQC) \cite{Bojowald:2006da} has been shown
to cure the big bang singularity and replace it with a big bounce
for Friedmann-Robertson-Walker (FRW) spacetimes with a massless
canonical scalar field \cite{BBLQC}. The bounce occurs when the
matter energy density reaches a Planckian value and quantum gravity
effects behave repulsively. The results have been generalized by
considering the approximate modifications to the Friedmann equation
implied by LQC. In particular, the models are non-singular for more
general forms of the scalar field potential \cite{Singh:2006im}, the
inclusion of anisotropies in the Bianchi I model \cite{DWKV}, the
Schwarzschild black hole interior \cite{BH}, and big rip models with
constant equation of state \cite{Sami:2006im}. These results provide
hope that maybe a general singularity resolution theory exists for
the full theory of loop quantum gravity (LQG), of which LQC is a
specialized model.

In this letter we study both canonical and phantom scalar fields
with exponential potentials using the effective Friedmann equations
of LQC. We find a class of singular cosmologies characterized by a
bounded Hubble parameter, but diverging time derivative of the
Hubble parameter occurring in finite proper time. While the energy
density is bounded, the pressure diverges, a type of singularity
known in general relativity as a sudden singularity
\cite{Barrow:2004xh}.

The energy density and pressure of the scalar field are,
respectively, given by
\begin{eqnarray}
\label{rho}
 \rho = \pm \frac{1}{2}\dot{\phi}^2+V
  \,, \qquad
 p = \pm \frac{1}{2}\dot{\phi}^2-V
  \,,
 \label{pressure}
\end{eqnarray}
where the plus sign indicates a canonical scalar field and the minus
corresponds to a phantom field. The local conservation of
energy-momentum leads to the Klein-Gordon equation governing the
dynamics of the scalar field
\begin{equation} \label{Klein-Gordon}
 \ddot{\phi} + 3 H \dot{\phi} \pm \frac{\partial V(\phi)}{\partial \phi} =
 0 \,,
\end{equation}
with $V(\phi)$ representing the scalar field potential. We will
consider exponential potentials of the form
\begin{eqnarray}
 \label{Vphi}
 V=V_0e^{\lambda\kappa\phi} \,,
\end{eqnarray}
where without loss of generality we take $\lambda > 0$.
Scalar fields with exponentials commonly arise in effective theories
derived from dimensional reductions of higher dimensional models,
and appear, for example, in the four-dimensional effective theory in
the ekpyrotic scenario \cite{Khoury:2001wf}.
We have found qualitatively similar results for simple polynomial
potentials.


In general relativity the Friedmann equations for a spatially flat,
isotropic universe are given by
\begin{eqnarray} \label{Friedman-GR-H}
 H^2 = \frac{\kappa^{2}}{3} \rho
 \,, \qquad
 \dot{H} = -\frac{\kappa^{2}}{2}  \left( \rho + p \right)
 \,,
 \label{Friedman-GR-Hdot}
\end{eqnarray}
where $H\equiv\dot{a}/a$ is the Hubble rate, $a$ the scale factor,
and $\kappa^2 = 8 \pi G_N$.

General relativistic and spatially flat cosmologies with canonical
scalar fields and exponential potentials are always singular. With a
positive potential
\cite{Halliwell:1986ja,Burd:1988ss,Copeland:1997et}, $V_0>0$, the
scale factor is monotonic with a big bang singularity where
$H\to\infty$ as $a\to0$ at a finite proper time in the past of an
expanding universe, or the time reverse (a big crunch singularity
where $H\to-\infty$ as $a\to0$ at a finite time in the future). With
a negative potential \cite{Heard:2002dr}, $V_0<0$, there exist
turning points where $H=0$ but these are always maxima of the scale
factor where $\dot{H}<0$. Solutions have either a big bang
singularity in the past or a big crunch singularity in the future,
or both. A phantom scalar field with a positive exponential
potential, $V_0>0$, can avoid a big bang singularity, but suffers a
big rip \cite{Caldwell:1999ew} singularity where $H\to\infty$ and
$a\to\infty$ at a finite proper time in the future (or the time
reverse in the past) \cite{UrenaLopez:2005zd,Hao:2003ww}. Note that
one cannot have a phantom scalar field with negative potential as
the total energy density must be non-negative in a spatially flat
Friedmann cosmology.

To incorporate the quantum effects due to LQC, we consider
effective Friedmann equations with corrections of the form \cite{BBLQC, Singh:2006im}
\begin{eqnarray}\label{Friedman-LQC-H}
 H^2 &=& \frac{\kappa^2 \rho}{3} \left[1-\frac{\rho}{\rho_c} \right]
 \,, \\
 \dot{H} &=& -\frac{\kappa^{2}\left(\rho+p \right)}{2} \left[1- 2 \frac{ \rho }{\rho_c}
 \right]
  \,.
\label{Friedman-LQC-Hdot}
 \end{eqnarray}
where $\rho_c$ is a constant parameter measuring the loop quantum
effects.  The value of $\rho_c$ is on the order of the Planck
density, $\kappa^{-4}$.
Note that these equations are approximations to the true
quantum dynamics. They have been validated from more
rigorous constructions of the quantum dynamics for simpler models with a
massless scalar field in \cite{BBLQC}. Some additional quantum state dependant
corrections are suggested in \cite{Bojowald:2007qu, Bojowald:2008pu}, but they have yet to be compared
to rigorous quantum constructions. Furthermore, the effective Friedmann equation
of \cite{Bojowald:2008pu} has been derived assuming the LQC effects occur in kinetic dominated
regimes, which will not be the case for our solutions.
 The classical GR limit can be recovered by
letting $\rho_c \rightarrow \infty$, whence it is easy to see that
the GR Friedmann equations \eqref{Friedman-GR-H} are recovered. The
loop effects do not modify the form of the scalar field equations
and thus the Klein-Gordon equation \eqref{Klein-Gordon} is
unchanged.

The prediction of a bounce in LQC with canonical scalar fields and
positive potentials can be understood from the modified Friedmann
equation \eqref{Friedman-LQC-H}. When the matter energy density
reaches the critical value $\rho_c$, the Hubble rate becomes zero
triggering the bounce. Therefore, in the case of the canonical
scalar field with positive exponential potential ($V_0 > 0$), the
past big bang or future big crunch singularity of the GR solutions
is replaced by a bounce and the cosmology becomes non-singular.

\begin{figure}[htbp]
  \begin{center}
 \includegraphics[width=7.2cm]{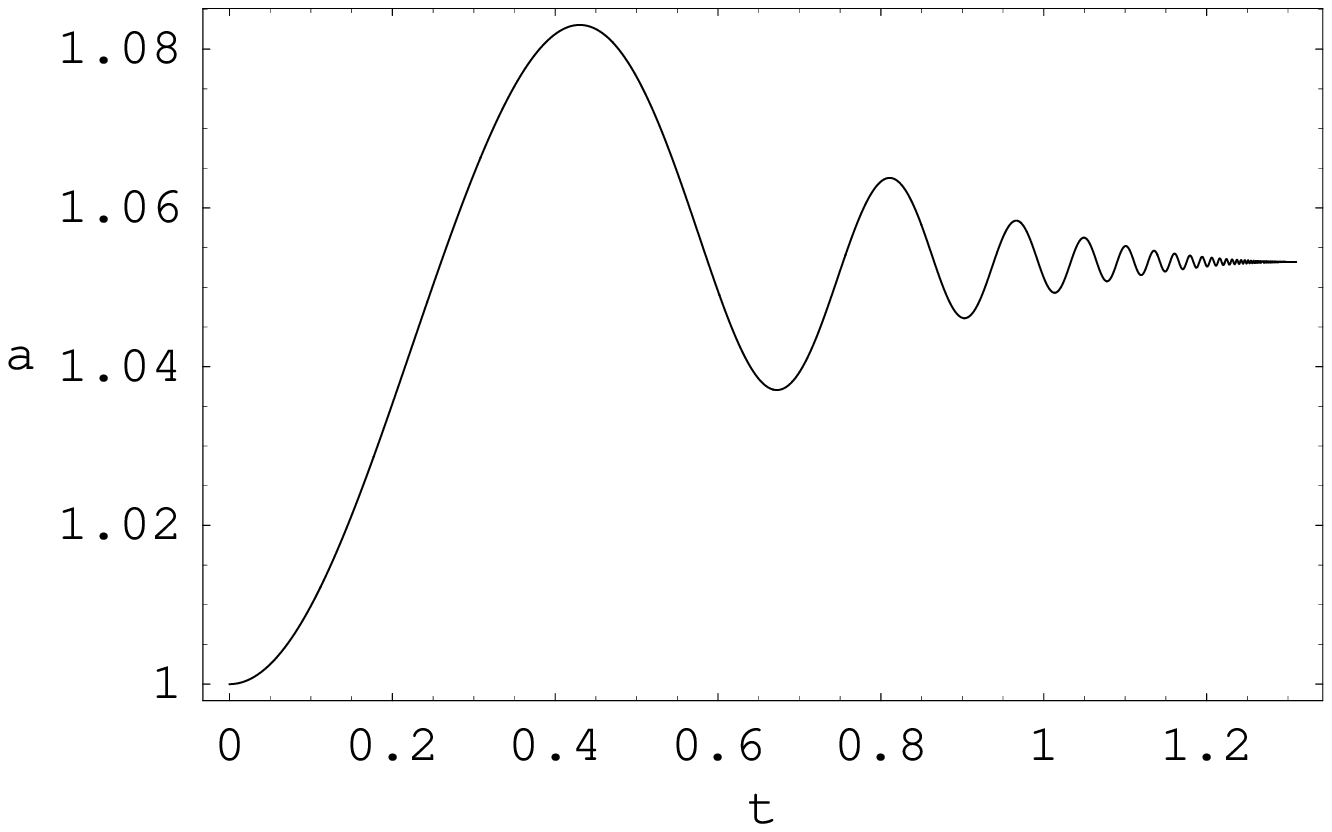}
 \includegraphics[width=7.2cm]{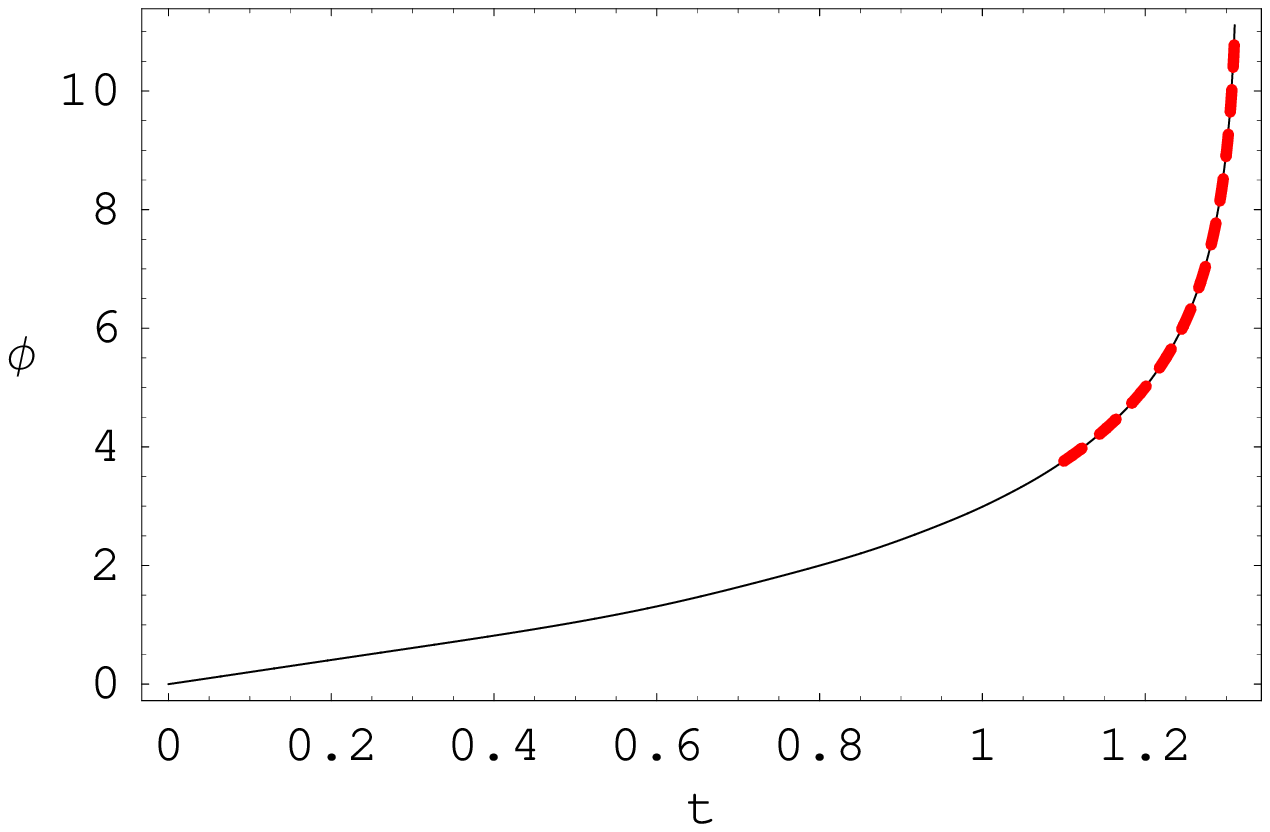}
 \includegraphics[width=7.2cm]{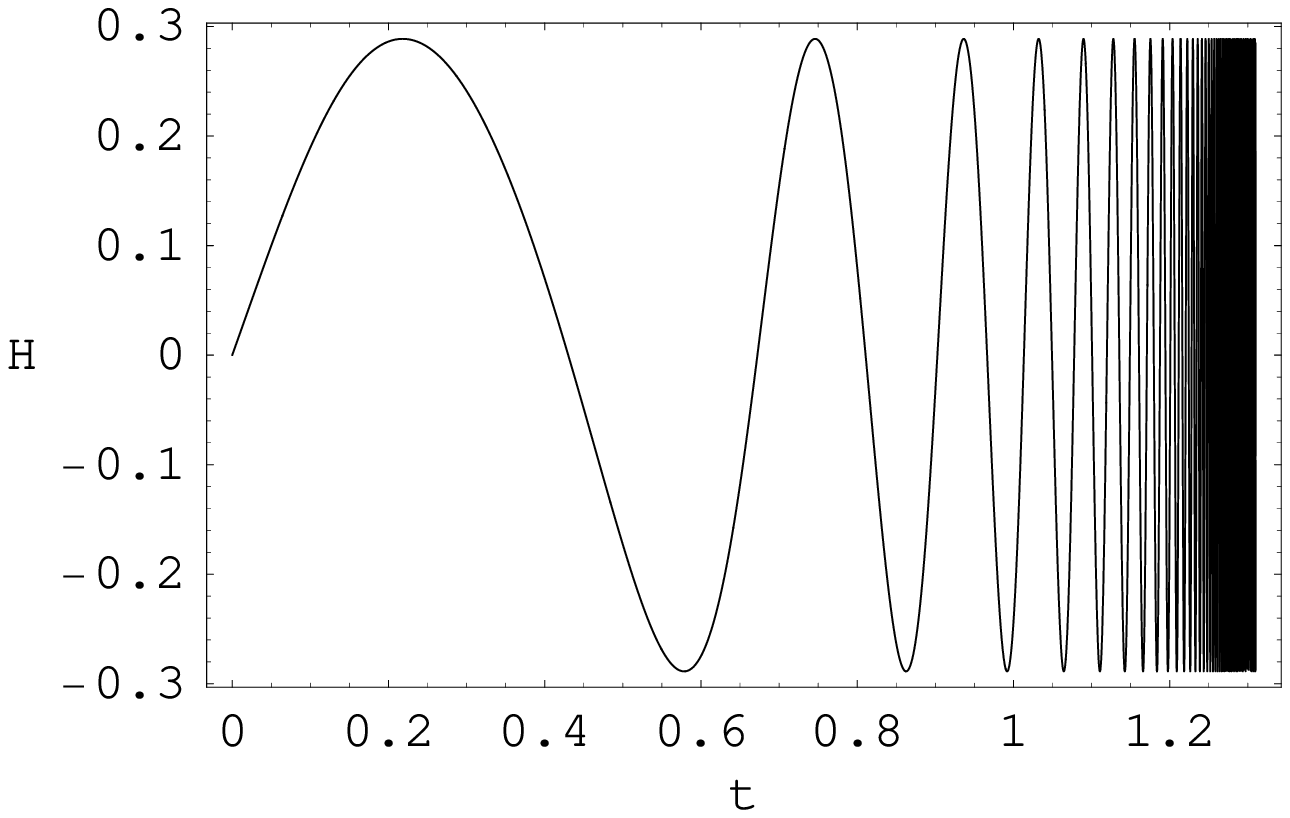}
 \includegraphics[width=7.2cm]{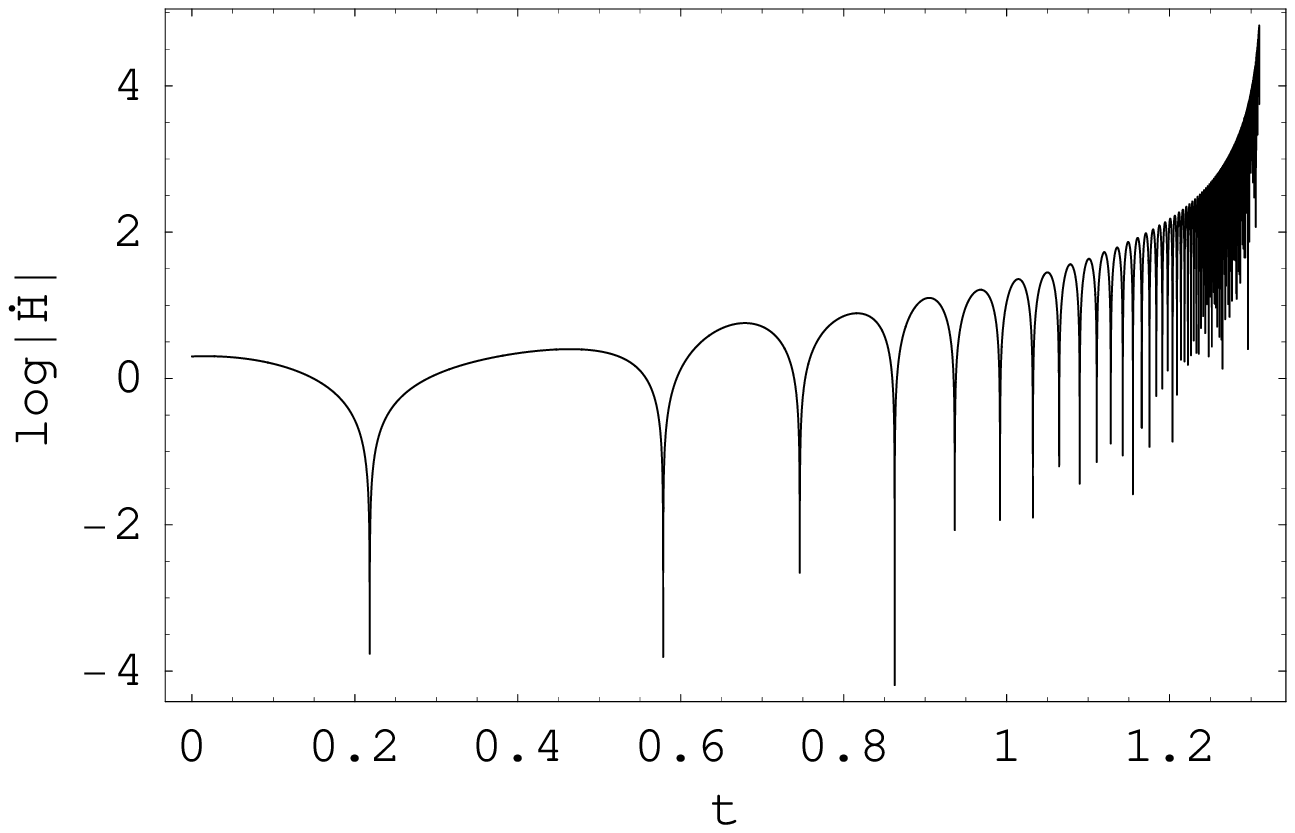}
 \end{center}
 \caption{Canonical scalar field with negative exponential potential.
Plots of $a$, $\phi$, $H$ and $\dot{H}$ as a function of time,
obtained from numerical simulations for $\kappa=1$, $V_0=-1$,
$\lambda=1$ and $\rho_c=1$. As initial conditions we set
$\phi(0)=0$, $\dot{\phi}(0)=\sqrt{2(\rho_c-V_0)}$, and $H(0)=0$.
The dashed curve in the plot of $\phi(t)$ is the prediction from the
approximate analytical solution \eqref{phi_aprox}.}
  \label{plot1}
\end{figure}

\begin{figure}[htbp]
 \begin{center}
\includegraphics[width=7.2cm]{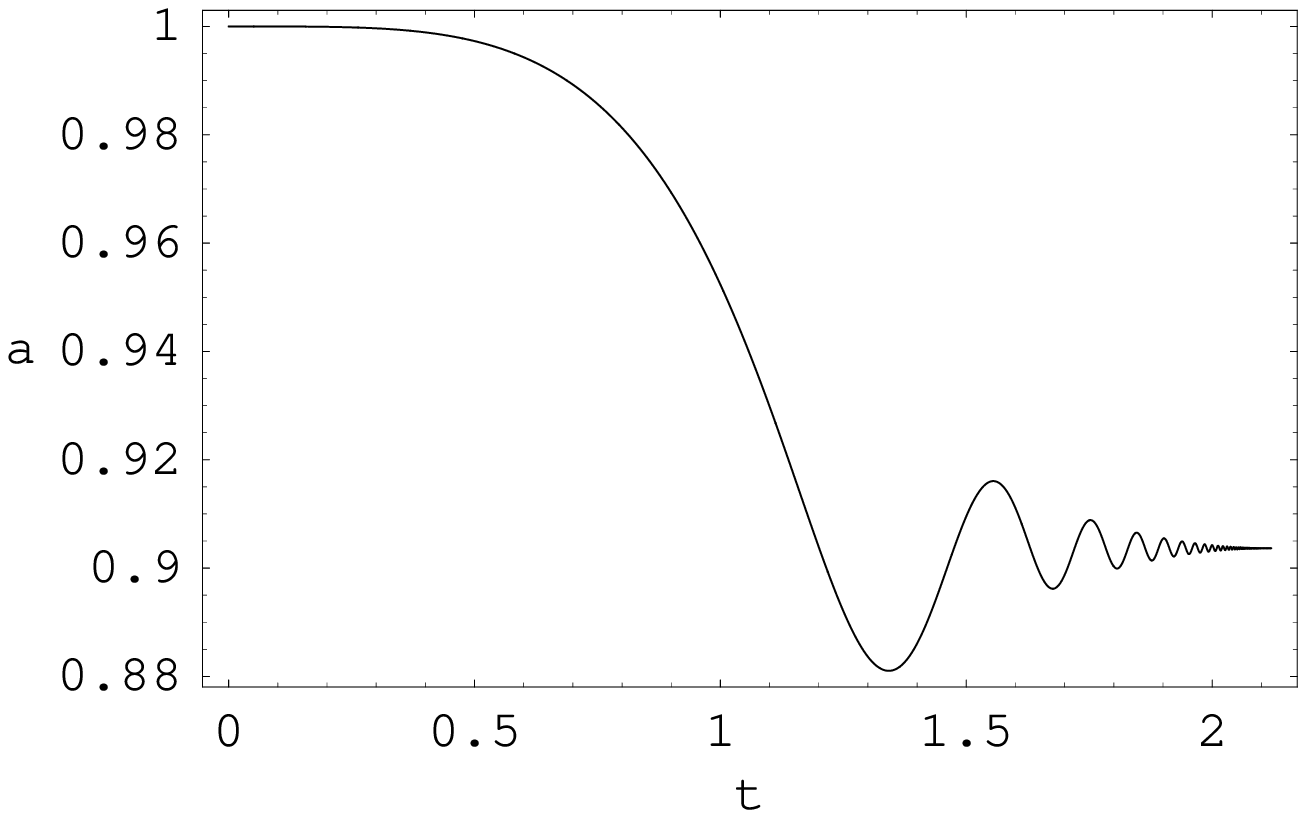}
\includegraphics[width=7.2cm]{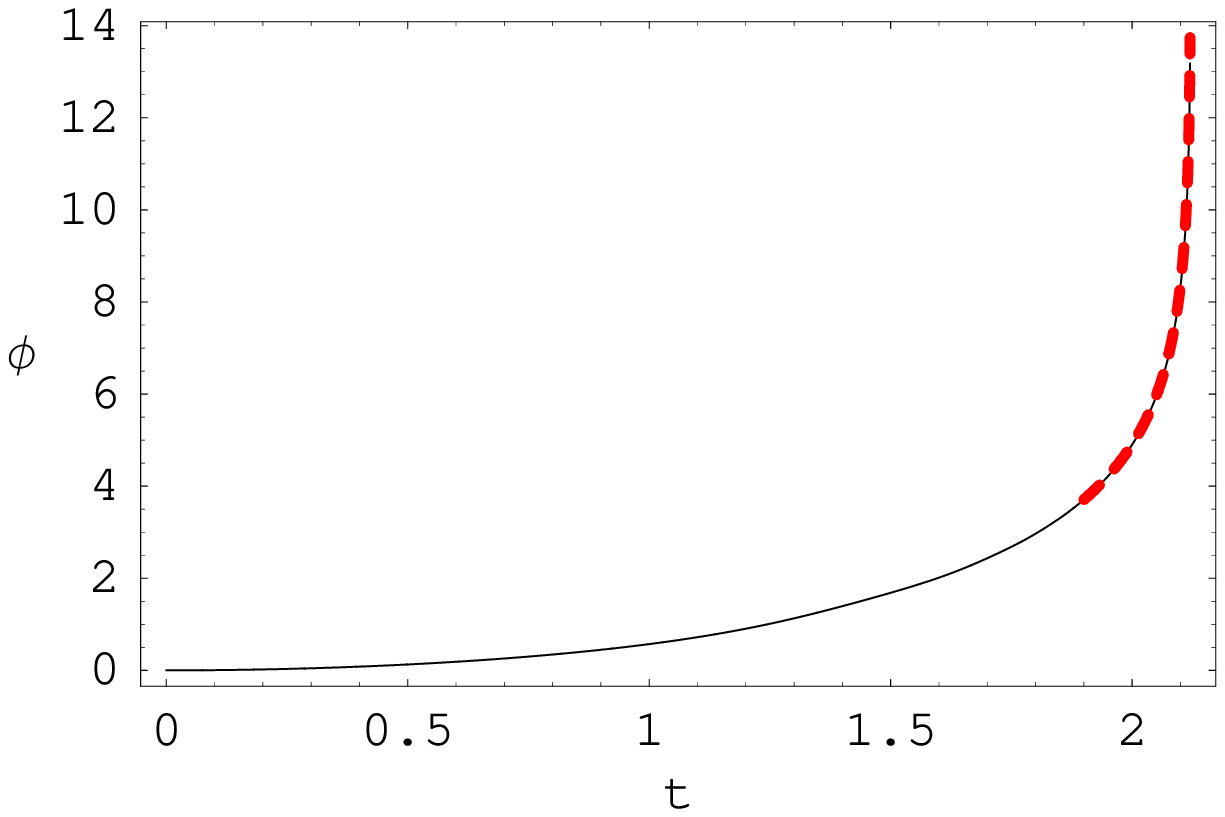}
\includegraphics[width=7.2cm]{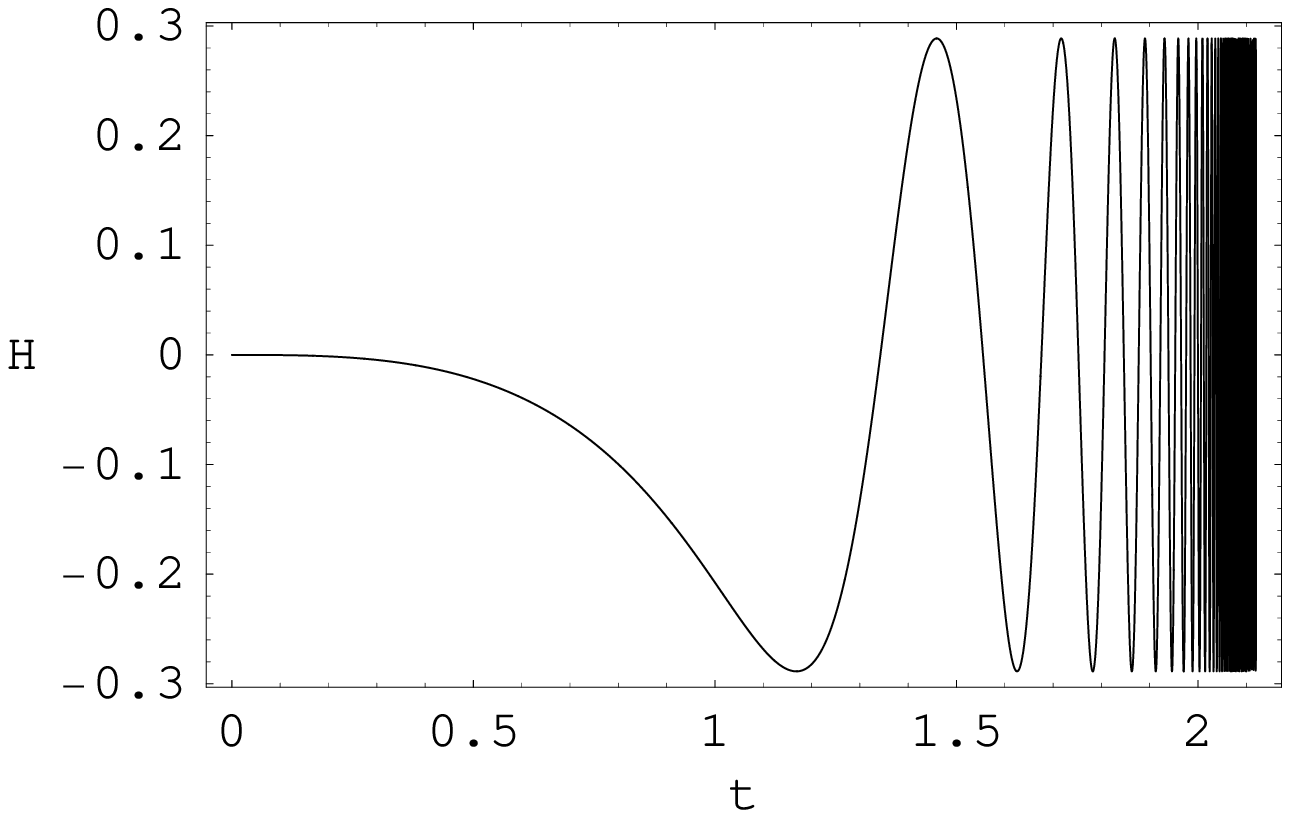}
\includegraphics[width=7.2cm]{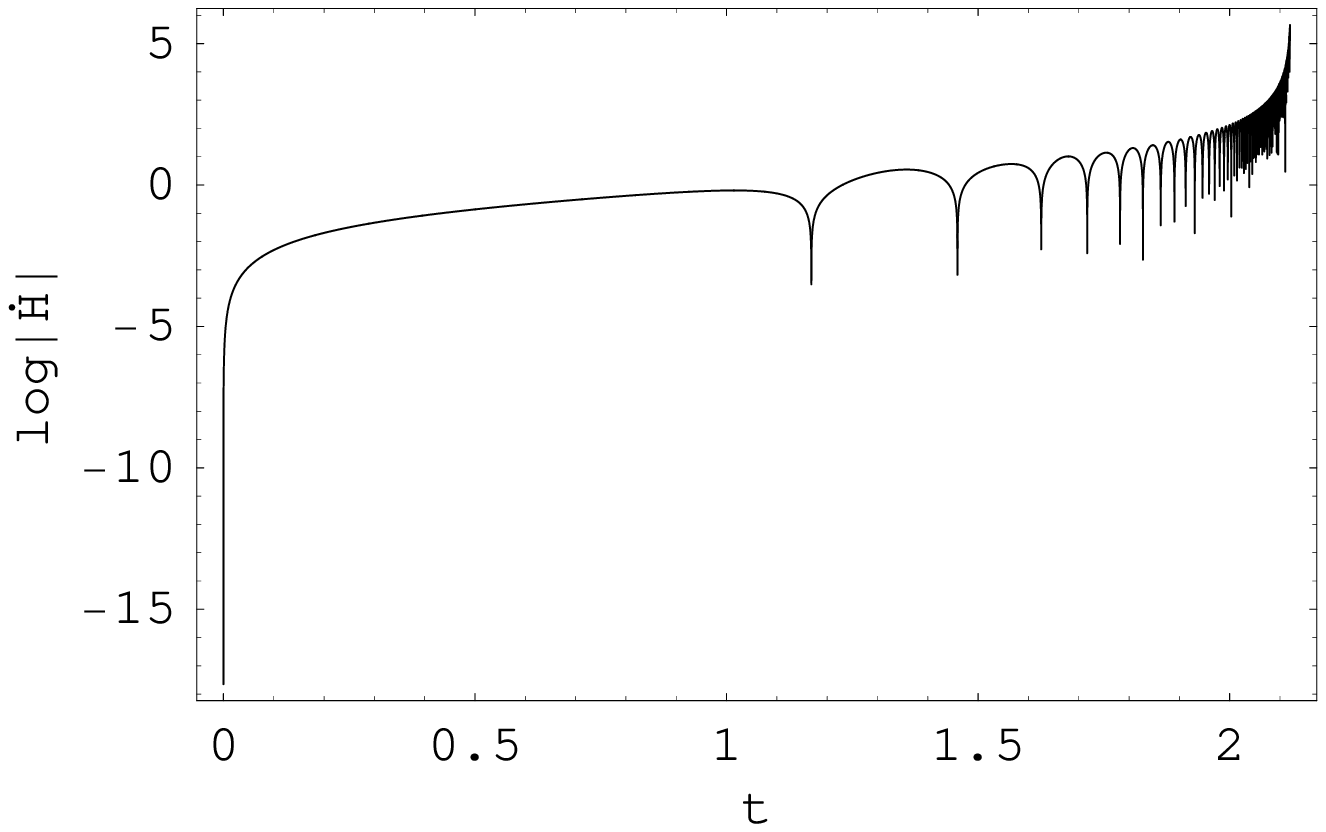}
    \end{center}
\caption{Phantom scalar field with positive exponential potential.
Plots of $a$, $\phi$, $H$ and $\dot{H}$ as a function of time,
obtained from numerical simulations, for $\kappa=1$, $V_0=1$,
$\lambda=1$ and $\rho_c=1$. As initial conditions we use
$\phi(0)=0$, $\dot{\phi}(0)=\sqrt{2(V_0-\rho_c)}$, and $H(0)=0$.
The dashed curve in the plot of $\phi(t)$ is the prediction from the
approximate analytical solution \eqref{phi_aprox}.}

 \label{plot2}
 \end{figure}

Numerical results for scalar field cosmologies with the loop quantum
equations of motion are shown in figure \ref{plot1} for the
canonical scalar field with negative potential and figure
\ref{plot2} for the phantom scalar. For the numerics we have taken
$\lambda=1$, $V_0=\mp\kappa^{-4}$, $\rho_c=\kappa^{-4}$ and set
$\kappa=1$. Initial conditions are chosen such that the scalar field
is zero and the universe is at a turn-around with $H(0)=0$ and
energy density equal to $\rho_c$. This implies a bouncing
(re-collapsing) phase for the canonical (phantom) case. The
qualitative results do not depend on the choice of initial
conditions.

Both the canonical and phantom scalar fields show similar behavior.
The solutions are characterized by repeated bouncing and collapsing
phases with the period of oscillations decreasing to zero in finite
proper time. As the period decreases, the expanding and collapsing
phases become shorter in duration and the scale factor approaches a
constant at the point when the evolution terminates. The scalar
field rolls down (up) the potential for the canonical (phantom) case
reaching infinity at some final value of t. The Hubble rate is
bounded, as expected from the LQC Friedmann equation
\eqref{Friedman-LQC-H}, while the time derivative of the Hubble rate
diverges. Thus the solutions have a ``quiescent''
\cite{Shtanov:2002ek} or ``sudden'' \cite{Barrow:2004xh} singularity
where $a$ and $H$ are both bounded but $\dot{H}$ diverges. The Ricci
curvature scalar, $R=6\dot{H}+12H^2$, diverges at a finite proper
time.\footnote{
We note that although the curvature diverges, these singularities
are weak in that they may be geodesically
complete~\cite{FernandezJambrina:2004yy,FernandezJambrina:2006hj}.}

We note that the case of a phantom scalar field with positive
exponential potential in LQC was previously studied in
\cite{Samart:2007xz}. Our numerical results are in qualitative
agreement, but we find that the solutions are in fact singular.

In order to verify that the singular nature of the solutions is not
a numerical artifact, we can find an approximate analytical solution
for the regime where $a(t)$ is approximately constant, which becomes
increasingly accurate as the singularity is approached. The total
density, $\rho$, remains bounded as the kinetic energy,
$\dot\phi^2/2$, and potential energy, $V(\phi)$, separately diverge.
Setting $\rho=0$ we can use the definition of the energy density
\eqref{rho} and solve for $\phi(t)$ to get
\begin{equation}\label{phi_aprox}
 \phi(t) = -\frac{2}{\lambda\kappa} \ln \left[e^{-\lambda\kappa\phi_0/2}-\lambda\kappa \sqrt{\frac{|V_0|}{2}} (t-t_0)
 \right] \,.
\end{equation}
Here $t_0$ refers to a reference time near the singularity where the
scale factor approaches its constant value and $\phi_0=\phi(t_0)$.
Note that this solution applies to both the canonical and phantom
scalar fields approaching the singularity. In addition, because
the energy density is bounded and the pressure grows, the oscillating solutions
occur in a regime where the kinetic energy of the field approximately equals the potential energy.
 The analytical solution
diverges at a critical value of $t$ given by
\begin{equation}
 t_c= t_0 + \frac{1}{\lambda\kappa}\sqrt{\frac{2}{|V_0|}} e^{-\lambda\kappa \phi_0/2}
 \,.
\end{equation}
Both $\phi$ and $\dot{\phi}$ diverge at this time, and we see that
also the pressure and $\dot{H}$ diverge, indicating a curvature
singularity at a finite time, in agreement with the numerical
results. A comparison of the numerical solution to the analytic
approximation shows that the two are in close agreement and the
approximation becomes better as the critical time is approached.
This can be seen in figures \ref{plot1} and \ref{plot2} where the
red dashed line is the analytic solution \eqref{phi_aprox} which
matches well the numerical solution approaching the singularity at
$t_c$. Thus the singularity is a genuine feature of the LQC
equations and not a numerical instability.

We have shown that the class of exponential potentials studied here
gives rise to a singularity in LQC despite the quantum corrections
in equation (\ref{Friedman-LQC-H}).
We can show that this is a consequence of the potential being
unbounded. The effective Friedmann equation (\ref{Friedman-LQC-H})
requires that $\rho$ is bounded, and for a canonical field, with
non-negative kinetic energy, the upper bound on $\rho\leq\rho_c$
also requires that $V\leq \rho_c$, but the potential is not bounded
from below by the effective Friedmann equation
(\ref{Friedman-LQC-H}). We can write the evolution equation
(\ref{Friedman-LQC-Hdot}) as
\begin{equation}
 \dot{H}
  = -\kappa^2 (\rho-V) \left[1- 2 \frac{\rho}{\rho_c} \right]
 \,,
\end{equation}
Thus for a canonical field with non-negative kinetic energy,
$\dot{H}$ remains bounded if $V(\phi)$ is bounded from below.
Conversely, $\dot{H}$ is unbounded only if $V(\phi)$ is unbounded
from below. Similarly for a phantom field with negative kinetic
energy one can show that $\dot{H}$ is unbounded only if $V(\phi)$ is
unbounded from above.

\begin{figure}[htbp]
 \begin{center}
\includegraphics[width=7.2cm]{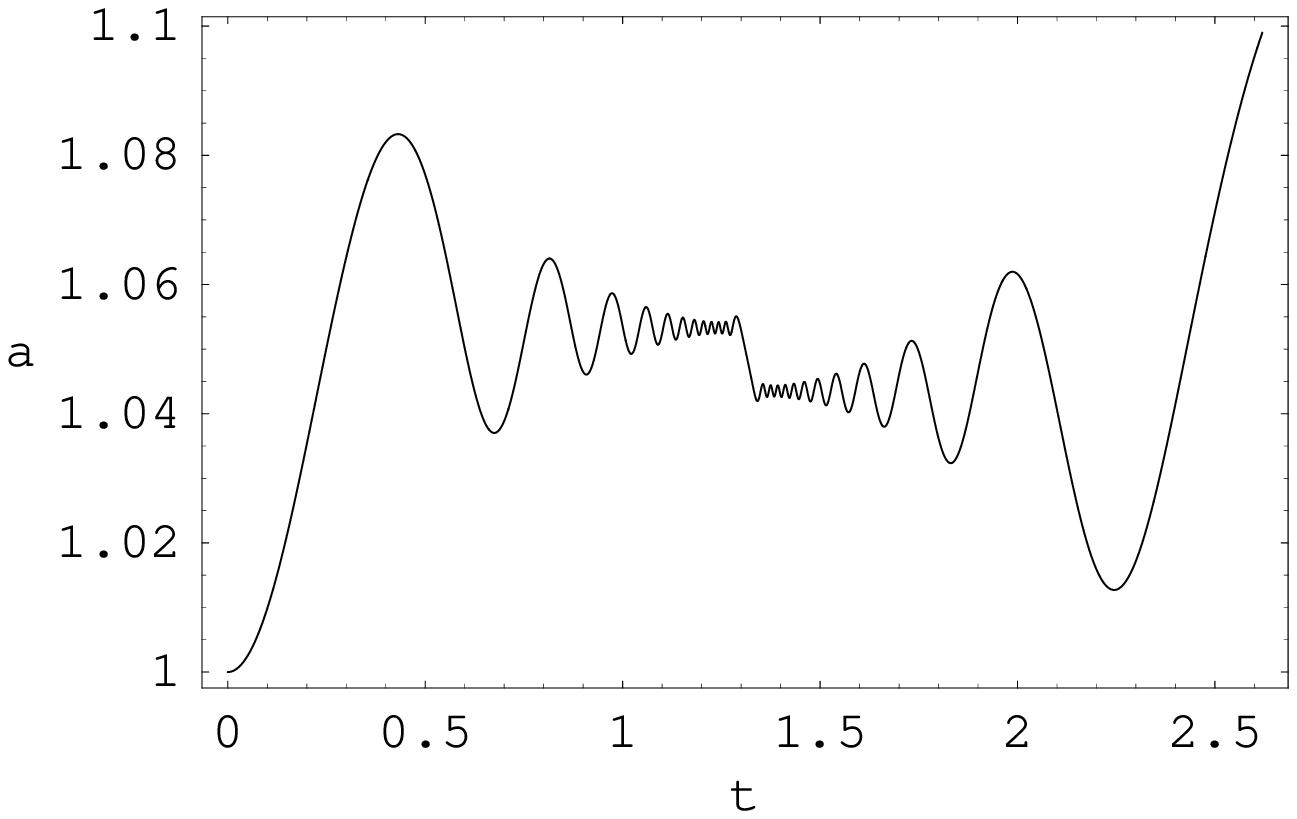}
\includegraphics[width=7.2cm]{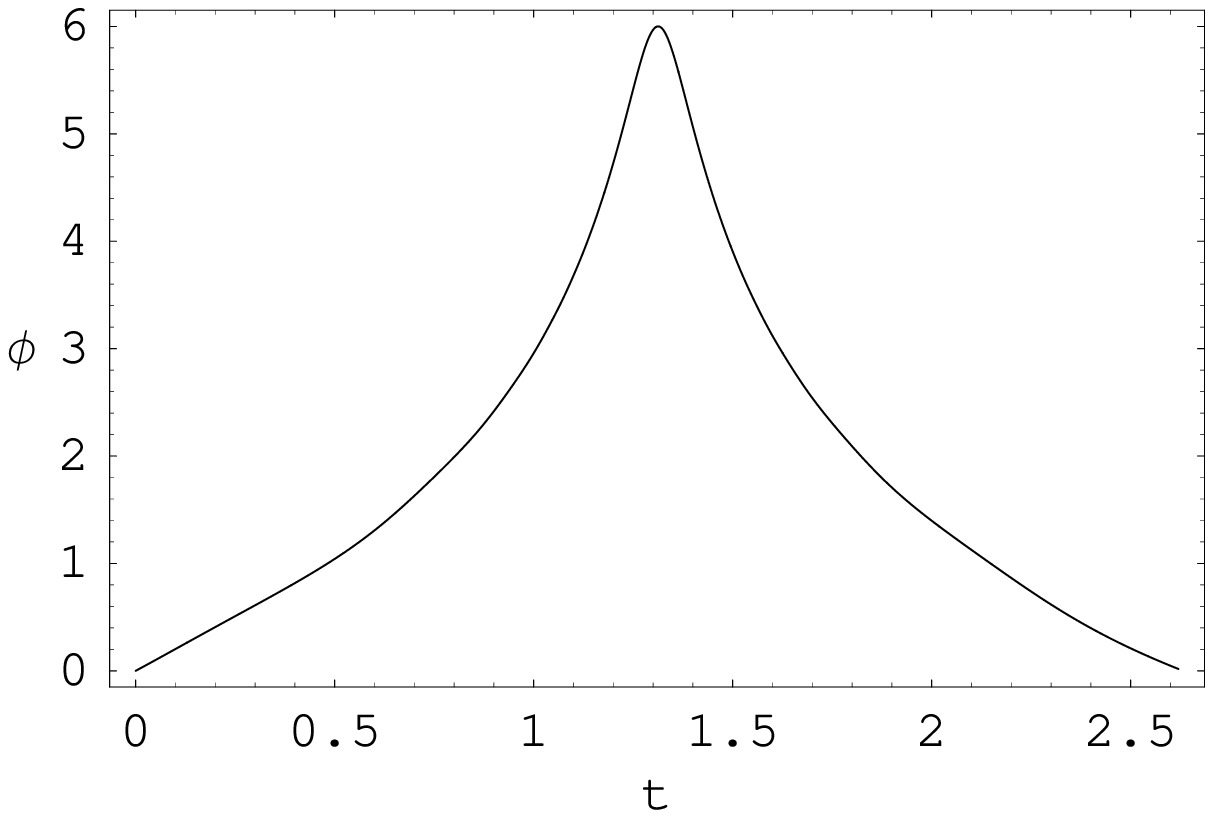}
\includegraphics[width=7.2cm]{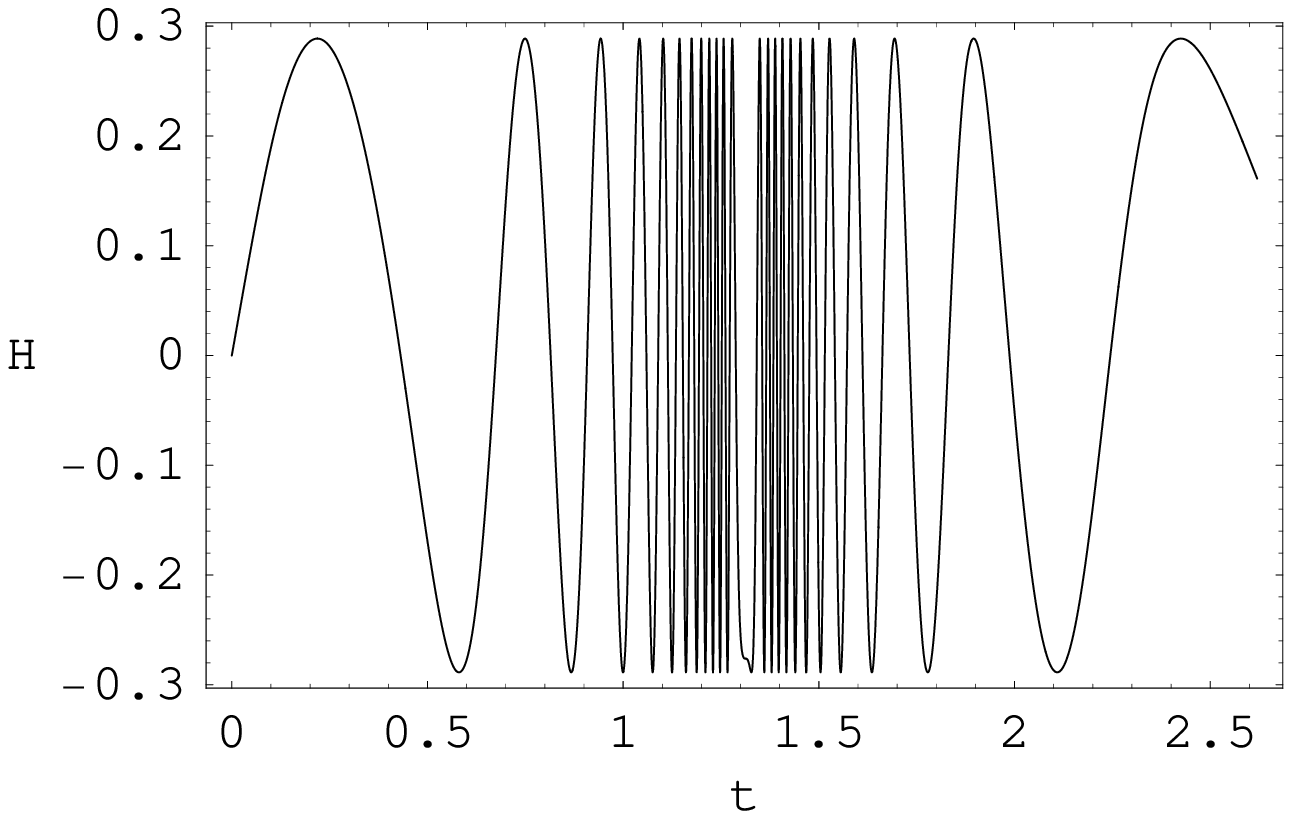}
\includegraphics[width=7.2cm]{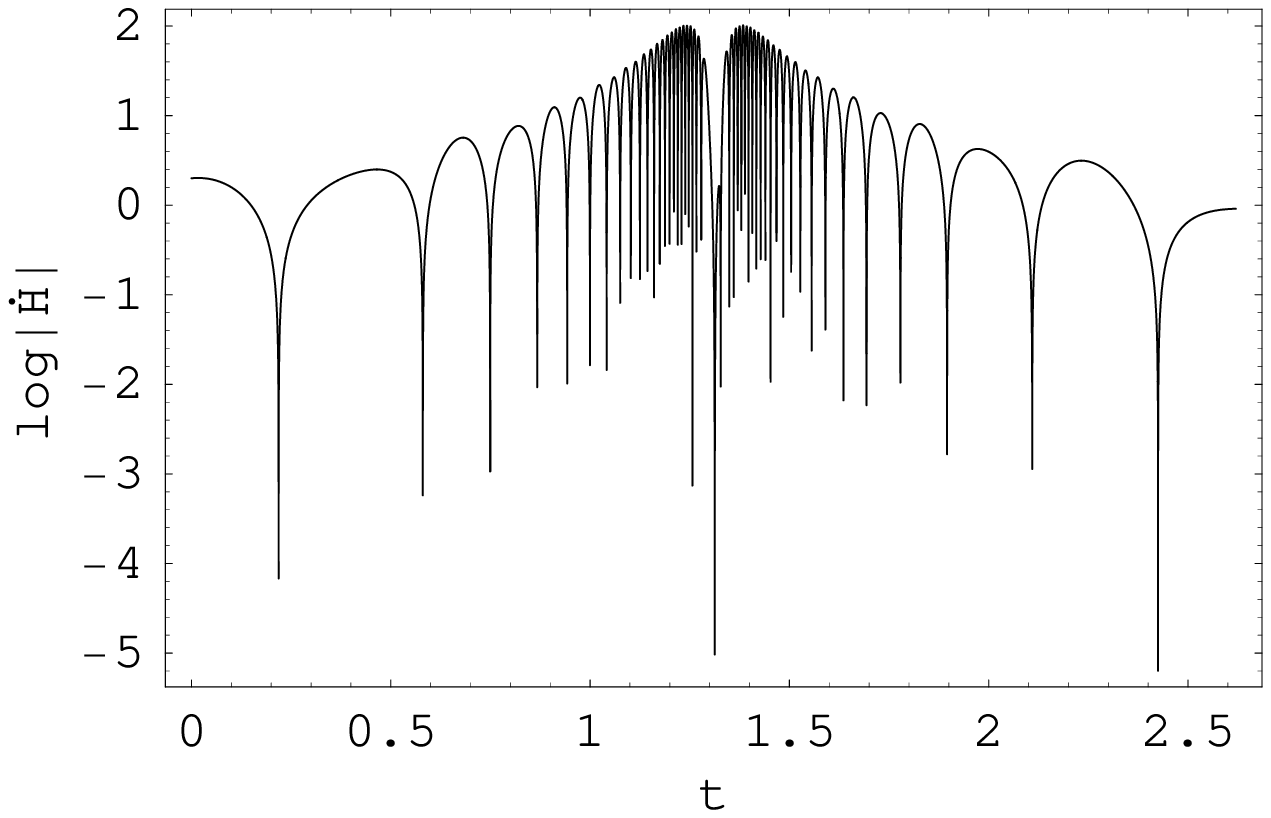}
    \end{center}
\caption{Canonical scalar field with bounded (double exponential)
potential. Plots of $a$, $\phi$, $H$ and $\dot{H}$ as a function of
time, obtained from our numerical simulations, for $\kappa=1$,
$V_0=-1$, $\lambda=1$, $V_2=e^{-6}$, $\lambda_2=2$. As initial
conditions we set $\phi(0)=0$, $\dot{\phi}(0)=\sqrt{2(\rho_c-V_0)}$,
and $H(0)=0$.
}
\label{plot3}
 \end{figure}

If we modify the exponential potential (\ref{Vphi}) to introduce a
lower bound, the sudden singularity disappears. In figure
\ref{plot3} we show numerical results for a canonical scalar field
with a potential of the form
\begin{equation}
 V(\phi) = V_0 e^{\lambda\kappa\phi} +V_2 e^{\lambda_2 \kappa\phi}
 \,,
\end{equation}
where $V_0 < 0$, $\lambda > 0$, as before, but $V_2 >0$ and
$\lambda_2 > \lambda$. These conditions ensure that the potential is
bounded from below and grows as a positive exponential for large
values of $\phi$. In the numerical run, the parameter values were
chosen to be $V_0=\kappa^{-4}$, $\lambda=1$,
$V_2=e^{-6}\kappa^{-4}$, and $\lambda_2=2$, and we have chosen the
same initial conditions as the single exponential case. With these
parameter values the potential minimum is located at
\begin{equation}
 \phi_{min} = \frac{1}{\kappa (\lambda_2-\lambda)} \ln \left( -\frac{V_0 \lambda}{V_2 \lambda_2} \right) \approx 5.3 \kappa^{-1}
 \,.
\end{equation}

The numerical run indicates that the solution behaves as in the
single exponential case with oscillatory expansion/collapse phases
with decreasing period until the potential minimum is reached, after
which the period of oscillations grows. The solution is non-singular
as is shown by the boundedness of both the Hubble rate $H$ and its
time derivative $\dot{H}$. The scalar field passes through the
potential minimum and rolls up the potential only to turn around at
a point where the potential is positive which is roughly given by
$\phi\approx 6\kappa^{-1}$ in figure \ref{plot3}. Similar behavior
occurs in the phantom case for a double exponential potential with
$V_0 >0$ and $V_2 < 0$ where the potential is now bounded from
above, and the behavior is also non-singular.

We have shown that singular solutions exist to the LQC effective
Friedmann equations for potentials that are not bounded from below
(above) for a canonical (phantom) scalar field. It is interesting
that sudden singularities, which may appear to be somewhat contrived
in general relativity \cite{Barrow:2004xh}, appear in quite simple
models in LQC. While this may indicate that generic singularity
resolution is not a feature of loop quantum gravity, it may simply
indicate that the effective equations of LQC break-down in some
specific cases. The scalar field Lagrangians which give rise to
singular behavior also may themselves be regarded as sufficiently
pathological that this need not be considered as a significant
limitation of LQC.\\
\\
\textbf{Acknowledgements:} TC would like to thank David Coule and Marco Bruni
for helpful discussions during the course of this work.
We thank Martin Bojowald, Leonardo Fern\'andez, Ruth Lazkoz and
Param Singh for useful comments.
AC is supported by FCT (Portugal) PhD fellowship SFRH/BD/19853/2004.
KV is supported by the Marie Curie Incoming International Grant
M1F1-CT-2006-022239. DW is supported by the STFC.

{}
\end{document}